
\magnification =\magstep1  
\nopagenumbers
\headline={\hss -- \folio\ -- \hss}
\hsize=6.0 truein
\vsize=8.5 truein
\voffset=0.25 truein
\hoffset=0.25 truein
\raggedbottom
\def\hangpara{\par\hangindent 25pt\noindent}
\def\Lya{Ly$\alpha$ }
\def\kms{km~s$^{-1}$ }
\def\etal{et al. }
\def\cm-2{cm$^{-2}$ }
\def\s-1{s$^{-1}$ }
\def\Hz-1{Hz$^{-1}$ }
\def\sr-1{sr$^{-1}$ }
%
%
%
%
\centerline{}
\vskip 1.0truein
\centerline{\bf UPPER LIMITS ON METALS IN QUASAR}
\centerline{\bf LYMAN-ALPHA FOREST CLOUDS: ABSENCE}
\centerline{{\bf OF C~IV LINES IN ECHELLE SPECTRA}~\footnote{\raise3pt\hbox{1}}
{Based on observations obtained at Lick Observatory, University of
California.}}
\bigskip
\centerline{ DAVID TYTLER\footnote{\raise3pt\hbox{2}}
{Department of Physics, University of California, San Diego}
{\raise3pt\hbox{,}}
\footnote{\raise3pt\hbox{3}}{Center for Astrophysics and Space Sciences,
0111, University of California,
San Diego, La Jolla, CA 92093--0111. (tytler@cass155.ucsd.edu,
fan@cass154.ucsd.edu)}
and XIAO-MING FAN{\raise3pt\hbox{3,}}
\footnote{\raise3pt\hbox{4}} {Dept. of Astronomy, Columbia University,
New York.}
}
\vskip 0.6truein
\centerline{To Appear in the Astrophysical Journal Letters}
\bigskip
\centerline{Received: \hbox to 1.5truein{$\underline{\rm December~13,~1993}$}}
\centerline{Accepted: \hbox to 1.5truein{$\underline{ \rm January~13,~1994}$}}
\vfill\eject
%
%
\centerline{ABSTRACT}
\bigskip

Recently Lu presented tentative evidence for C~IV lines in QSO \Lya
forest systems with strong lines. We have  performed a similar search
for C~IV in our 10~\kms echelle spectra of the bright QSO HS~1946+7658.
We shifted the spectra to align the expected positions of the C~IV lines
in 65 \Lya systems, then added them.  The resulting composite spectrum,
equivalent to 390 hours of exposure time on the Lick 3-m telescope, has
a signal-to-noise ratio of 80 per 0.025~\AA\ in the rest frame of the
absorbers.  We do not see any C~IV lines down to a $2\sigma$ limit of
W$(1548) \leq 1.4 $~m\AA, about one-fifth of the strength of the lines
seen by Lu. The C~IV lines which Lu saw must be restricted to rare \Lya
systems with
large H~I column densities $\geq 10^{14}$~cm$^{-2}$, which are too rare to
show C~IV in our sample.  More common \Lya systems with H~I column densities
of $10^{13}$ -- $10^{14}$~cm$^{-2}$ do not show C~IV lines.
If their ionization is H/H~I $= 10^4$ then they have [C/H] $\leq -2.0$.

\bigskip
\noindent
{\it Subject Headings:} galaxies: abundances -- intergalactic medium
-- quasars: absorption lines

\vfill\eject
%
%
\bigskip
\centerline{1. INTRODUCTION}
\medskip

Quasar Lyman-alpha (\Lya) forest absorption systems do not show metal lines in
spectra with 1~\AA\ resolution (e.g., Sargent et al. 1980; SYBT), but
there has been much discussion of the possibility that they do contain
metals (e.g., Tytler 1987; Verner, Tytler \& Barthel 1994). Our  recent
discovery (Lanzetta \etal 1994) that at least $35\pm 10$\%, and likely
$0.65 \pm 0.18$\% of \Lya clouds at $z_{abs} \leq 1$ arise from the outer
regions of luminous galaxies confirms our earlier proposition that \Lya
forest systems are related to metal-line systems (the ``one population''
concept discussed by Tytler 1987), and makes us expect to find metals in at
least some of these low-redshift systems.

We need three pieces of information to derive an observational limit
on metal abundances:
(1) a limit on the strength of the metal absorption lines,
(2) a measurement of the H~I column density $N$(H~I), and
(3) an estimate of the level of ionization, which is used to calculate
the abundance of the elements from the observed ions.

In this paper, we give improved upper limits on the strength of C~IV lines
in \Lya systems. There are two ways to obtain such limits:  one is to observe
the systems with  the highest $N$(H~I); the second is to
form an average spectrum at the expected positions of the metal lines
observed in many \Lya systems.

The first method relies on the fact that metals should be easier
to detect in systems with the largest total column densities.
Sargent \& Boksenberg (1983) and Chaffee et al. (1985, 1986) found limits
of [M/H] $\leq -3.5$ for two \Lya clouds with large H~I column densities of
$10^{16} \leq N{\rm (H~I)} \leq 10^{17}$cm$^{-2}$.
There are two problems with this method:
first, these \Lya clouds are very rare because their $N$(H~I)
is 300 times larger than those normally observed;
second, we do not know if they
actually have the largest total column densities because we do not know their
ionizations.

We use the second method to look for metals in common \Lya clouds
which have low $N$(H~I).  We shift spectra to align the expected
positions of the metal lines, then we sum them to increase the SNR
(signal-to-noise ratio).
To shift a spectrum to the frame of a \Lya cloud at redshift
$z_i$ we divide the observed wavelengths by $(1+z_i)$, and we do this
for each \Lya cloud which could have a metal line in the spectrum.
This {\it shift and sum} technique was introduced by
Norris, Hartwick \& Peterson (1983), and was used by Lu (1991),
who claimed a tentative detection of C~IV at the 99.99\% significance
level. Here we use the same method on better data. Our echelle
spectra yield more sensitive limits because:
(1) the expected positions of the metal lines are more accurately
located because we have higher resolution and more accurate wavelengths,
(2) the continuum level is better determined,
(3) we get more reliable identifications of weak lines in the C~IV region, and
(4) we know the $N$(H~I) for each \Lya system in our sample.
Echelle spectra do not
always give better sensitivity, because this also depends on the SNR per
\AA ngstrom, the number of \Lya lines, and their $N$(H~I) and ionization.

We consider only C~IV because its doublet lines at 1548.20 and 1550.77~\AA\
are likely to be the strongest metal lines at rest frame wavelengths
greater than 1200~\AA, where we have good data.

\bigskip
\centerline{2. LIMITS ON THE STRENGTH OF C~IV IN TYPICAL LYMAN-$\alpha$
SYSTEMS}
\medskip

We use spectra of only one QSO, HS1946+7658 (Hagen \etal 1992),
which is the most luminous known QSO with $V=15.85$ and $z_{em} = 3.051$.
In August 1992 and June 1993, we obtained spectra with the Hamilton echelle
spectrograph on the Lick 3-m telescope. These spectra have 10 \kms resolution
and  cover 4220 -- 7251~\AA\ with inter-order gaps of 3.6~\AA\ at
the blue end and 42.6~\AA\ at the red end.
The SNR is one (per 0.1~\AA\ pixel in the observed frame)
at the blue end, 16 in the \Lya emission line, and 7 at the red end, but
we also have complete wavelength coverage with intermediate resolution
spectra in optical with high SNR, so our line identifications are unusually
complete and accurate.  Fan \& Tytler (1994) discuss the properties of the
\Lya forest clouds and the abundances of the metals in the damped \Lya
system in this spectrum.

We rebinned all the spectra into new pixels with rest frame width
$\Delta \lambda = 0.025$~\AA, which is 4.8 \kms at 1548~\AA,
close to the original mean pixel width of 0.026~\AA\ .
Fan \& Tytler (1994) have already normalized the continuum of this spectrum,
and for this project we replace all absorption
lines that they listed in the C~IV region with noisy continuum with the
same SNR as the adjacent continuum.

For the shift and sum method, we need high SNR for both the \Lya and
the C~IV lines, so we consider systems with $z_{abs}= 2.471 - 3.051$ that
have \Lya lines at 4220 -- 4925~\AA\ and C~IV lines at 5374 -- 6282~\AA\ and
where the spectra have SNR=10--14 per 0.10~\AA\ pixel (observed frame)
from 6 hours of exposure time. We shift the spectra of 65 \Lya systems
to align the expected positions of the C~IV lines, then we add the spectra.
Because of the inter-order gaps, the number of spectral segments included
in the composite depends on the wavelength.
We have a maximum of 65 at C~IV~$\lambda 1548$, 58 at C~IV~$\lambda  1550$,
38 at the blue end, and 27 at the red end of the spectrum which we show in
Figure 1.
The resulting composite spectrum has an effective exposure time of 390
hours on the 3-meter telescope.

Figure 1 shows two composite spectra. The first, in panel (a),
is an unweighted average with SNR=80 per 4.8~\kms.  The second, in
panel (b), is weighted by $N$(H~I) with SNR=50.  The C~IV lines should
extend over at least five pixels because the original spectra had a FWHM
of two pixels.  Panels (c) and (d) in Figure 1 show the rest frame
equivalent width
for lines that are five pixels wide. There are no significant C~IV absorption
lines. The rest frame equivalent width for the 1548 line, which should be
twice as strong as the 1550 line, is $W_r= 1.4$~m\AA\ in the unweighted
and $2.6$~m\AA\ in the weighted composites.  When we use log~$N$(H~I) as an
alternative weight, the result is nearly identical to the unweighted
composite in panel (a) because all \Lya lines have very similar log~$N$(H~I).

The C~IV lines could be stronger than the $W_r$ values shown in panels
(c) and (d) because noise can reduce as well as increase $W_r$.  The
statistical limits on $W_r$ depend
on the SNR per pixel and the number of pixels which the lines are
expected to cover ($M_L$).  The $U \sigma $ limit is
$$
W_{r, max}= {U ( M_L^2/M_c+M_L )^{0.5}  \over (SNR^2 + U^2)^{0.5}}
\Delta \lambda  ~~\rm (\AA), \eqno(1)
$$
where $M_c$ is the number of pixels used to determine the continuum level.
When the continuum is well-determined so that  $M_c \gg M_L$
and the lines are weak so that $SNR \gg U$, this reduces to
$$
W_{r, max} \simeq {U M_L^{0.5} \over SNR} \Delta \lambda  ~~\rm (\AA), \eqno(2)
$$
which is appropriate for our data. We use $U=2$
because we know  where to expect the lines, and $M_L=5$, which gives
$2\sigma$ limits of 1.4~m\AA\ for the unweighted and 2.2~m\AA\ for
the weighted composites.  There are small dips at the positions of the
1548~\AA\ lines in both composites, but neither exceeds the
$2\sigma$ level.  The actual limits may be slightly higher because we
have ignored the small increase in SNR which accompanies the smoothing
which happens when data are rebinned, but this should be minimal because
we kept a roughly constant pixel size (we did not oversample).

\bigskip
\centerline{3. WHY DO OUR RESULTS DIFFER FROM LU'S?}
\medskip

Why did Lu (1991) see C~IV 1548 lines that are five times stronger
than our $2\sigma $ upper limit? He measured $W_r$ values
of 2.4, 7.1, 7.2, 9.2, 11.1, 6.1, 9.5 and 14.1~m\AA\ in his eight
composite spectra, which were all subsamples for \Lya lines of
increasing strengths, and he used thorough simulations to show that
most of these lines are statistically significant. For example, the
1548 line with $W_r =7.1$~m\AA\  in his sample with
$W_r (1216) \geq 0.4$~\AA\ is significant at the 99.99\% level because
no similar lines were seen in 10,000 simulations. We accept that his
lines are real and conclude that he saw them because his sample
includes strong \Lya lines, some of which have weak C~IV lines.
We do not see C~IV lines because our sample does not have any strong \Lya
lines and is comprised of common \Lya clouds that do not have C~IV lines
with  $W_r \geq 1.4$~m\AA.

Lu's sample is larger but shallower than ours.  We have 65 \Lya lines
from one QSO, but Lu used up to 324 lines from 14 QSOs. Our sample has
mean $N$(H~I)=$1.0 \times 10^{14}$ cm$^{-2}$ for 1548 and
$0.95 \times 10^{14}$ cm$^{-2}$ for 1550, but his subsamples have means
of at least $2 \times 10^{15}$ cm$^{-2}$, with large uncertainties.
Our means are 25 times lower because we detect much weaker \Lya lines, and
because we do not have enough coverage to see the rare systems with
large $N$(H~I).  Redshift is not a factor because our mean $z_{abs}=2.79$
is similar to his (2.4 -- 2.6).

Lu's sample is more likely to contain \Lya clouds with weak C~IV because:
(1) most systems with strong \Lya lines have C~IV (Tytler 1987),
(2) Meyer \& York (1987) showed that some \Lya systems have C~IV lines
which are too weak to be seen in the individual spectra used by Lu
(e.g. three of the four systems seen by Meyer \& York in Q2126-158
were not identified by Sargent, Boksenberg \& Steidel 1988), and
(3) we have identified and removed C~IV lines which are as weak as those
seen by Meyer \& York.
Lu recognized that his 1548 line could have resulted from 10 -- 20
out of 100 -- 200
\Lya clouds with $W_r (1548) \simeq 84$~m\AA\ each, but since his composite
spectrum remained similar as he raised $W_r (1216)$ above 0.4\AA\ ,
he suggested that all \Lya lines with $W_r (1216) \geq 0.4$ \AA\ have
about the same $W_r (1548)$.  This $W_r (1216) \geq 0.4$~\AA\ corresponds
to $N$(H~I)$\geq 10^{14}$\cm-2 for all $b \leq 70$\kms, and only 15 of our
65 \Lya systems have such high $N$(H~I), so our result is compatible with his.

\bigskip
\centerline{4. METAL ABUNDANCES OF THE \Lya CLOUDS}
\medskip

Our limit of $W_r \leq 1.4$~m\AA\ for a typical C~IV 1548 line corresponds to
$$\rm N(C~IV) \leq 3.5 \times 10^{11} { W_r \over 1.4~m\AA\ } ~(cm^{-2})
 \eqno(3)$$
because the line will be on the linear part of the curve of growth.
The corresponding abundance limit depends on the unknown level of ionization.
If we follow
SYBT, Chaffee \etal  (1986), and Lu (1991), and if we assume that the
clouds are highly ionized with
H/H~I$ = 10^4$ and C/C~IV = 5, then our typical \Lya clouds with
$N$(H~I)=$4 \times 10^{13}$~cm$^{-2}$
would have C/H $\leq 4.4 \times 10^{-6}$, which
corresponds to [C/H] $ \leq -2.0$ since solar C/H $=4.7 \times 10^{-4}$.
This cloud would be 16.7~kpc thick if it had constant density and
it was ionized by the Bechtold et al (1987) ''medium'' spectrum
of ionizing radiation, normalized to a flux of
$10^{-21}$ ergs \cm-2 \s-1 \Hz-1 \sr-1 at 912\AA\ at $z_{abs}=2.7$.
If all our \Lya clouds have similar ionization and C/H,
we should use the mean $N$(H~I)
instead of the typical value, which gives [C/H] $\leq -2.4$.
We can then use the composite spectrum weighted by $N$(H~I),
which gives [C/H] $\leq -2.2$.

Lu obtained lower abundances of [C/H] $\sim -3.2$ because his sample had clouds
with larger $N$(H~I), which he assumed had larger total column densities.
We do not know the ionization or total column density of any individual
\Lya system in our sample, nor do we know if the total column densities are the
highest at low, intermediate, or high $N$(H~I).  We think that it is
unlikely that the clouds all have the same gas density and ionization
because this would
imply too large a range in size (Sherwin 1984), so we do not expect a
simple one-to-one correspondence between $N$(H~I) and total column density,
but rather a large range of ionization at a given $N$(H~I).
Then if some high $N$(H~I) systems have high ionizations, they will
have the largest total column densities, and the absence of metals in
their spectra will give the best abundance limits of any individual system.
But we do not believe that all systems with high $N$(H~I)  give the best
individual abundance limits.

Although we have shown that common
\Lya clouds do not have C~IV lines of the strength seen by Lu, they could still
have abundances similar to the [C/H] $\simeq -3.2$ which he found for
clouds with high $N$(H~I), and they could have abundances similar to those
in the outer halo, which might be expected if \Lya clouds at high redshift
are associated with galaxies, as are most at $z \leq 1$ (Lanzetta et al 1994).

We will obtain much better abundance limits in the coming year for several
reasons.  HST spectra will distinguish high from low ionization,
although they will not give a quantitative measure of the ionization.
SYBT predicted that high-ionization \Lya clouds should have He~II
lines that are stronger than H~I, but no detectable He~I, whereas
low-ionization gas could show both He~I and He~II (see Fig. 3 of
Chaffee \etal 1986). Reimers \& Vogel (1993)
report that new HST spectra should have the sensitivity to detect He~I in
low-ionization clouds. However, two other expected new measurements
will not give the ionization. HST spectra of He~II $\lambda 304 $
will not give the ionization because He~II  has
the same electronic structure as H~I, so He~II/H~I
remains constant over a wide range of ionization and
is a measure of the ratio of the flux at the two ionization edges, not of
the ionization. Also, the potential detection of H$\alpha$ emission from
\Lya clouds will not give the ionization because the flux depends on the
product of $N$(H~I) and the flux of ionizing photons, not on the
density or ionization (Hogan \& Weymann 1987; Williams \& Schommer 1994).

The first few spectra from the  HIRES
echelle on the Keck telescope show that we should get SNR=120 per 0.1~\AA\
at 5000 -- 7000~\AA\  on a $V=15.87$ QSO in six hours.
This SNR is ten times better than that of the data used here -- 100 times more
photons
--  because Keck is ten times larger and because QSO spectra from the
Hamilton echelle on the Lick 3-m are degraded by dark current and readout
noise. Keck will be able to detect [C/H] ten times lower than the limits
of $-2.0$ to $-2.4$ which we obtained here.

We will also target \Lya clouds with larger $N$(H~I) and look at more QSOs,
and we will use HST and Keck to search for metal lines in the rest-frame
far UV.  Chaffee \etal (1986)
showed that C~III$~\lambda 997$ should be stronger than C~IV in low-ionization
systems, while O~VI~$\lambda\lambda $ 1031, 1037 should be stronger at high
ionization. Verner, Tytler \& Barthel (1994) point out that
Ne~VIII~$\lambda\lambda $ 770, 780 and Mg~X~$\lambda\lambda $609, 625
will be strong in very highly ionized gas.

\bigskip\bigskip

We thank Limin Lu for detailed comments,
Art Wolfe for showing us how he uses the spectrum of equivalent
widths, and Abe Oren for keeping the computers going.
FXM and DT were supported in part by NASA grant NAGW-2119
and by GO-3801.01-91A and GO-5492 from the Space Telescope
Science Institute, which
is operated by AURA, Inc. under NASA contract NAS5-26555.

\vfill\eject
%
%
\centerline{REFERENCES}
\bigskip

\hangpara Bechtold, J., Weymann, R. J., Zou, L., \& Malkan, M. A. 1987,
          ApJ, 281, 76
\hangpara Chaffee, Jr., F. H., Foltz, C. B., Bechtold, J., Weymann, R. J.
          1986, ApJ, 301, 116
\hangpara Chaffee, Jr., F. H., Foltz, C. B., R\"oser, H.-J.,
          Weymann, R. J., Latham, D. W. 1985, ApJ, 292, 362
\hangpara Fan, X.-M., \& Tytler, D. 1994, submitted to ApJ.
\hangpara Hagen, H.-J., Cordis, L., Engels, D., Groote, D., Haug, U.,
          Heber, U., K\"ohler, Th., Wisotzki, L., \& Reimers, D. 1992,
          A\&A, 253, L5
\hangpara Hogan, C. J., \& Weymann, R. J. 1987, MNRAS, 225, 1p
\hangpara Lanzetta, K. M., Bowen, D. V., Tytler, D. \& Webb, J. K. 1994
          submitted to ApJ.
\hangpara Lu, L. 1991, ApJ, 379, 99
\hangpara Meyer, D. M. \& York, D.G. 1987, ApJ, 315, L5
\hangpara Norris, J., Hartwick, F. D. A., \& Peterson, B. A. 1983,
          ApJ, 273, 450
\hangpara Reimers, \& Vogel, S. 1993, AApL, 276, L13
\hangpara Tytler, D. 1987, ApJ, 321, 49
\hangpara Sherman, R. D. 1984, ApJ, 284, 457
\hangpara Sargent, W. L. W., Young, P. J., Boksenberg, A., \& Tytler, D.
          1980, ApJS, 42, 41 (SYBT)
\hangpara Sargent, W. L. W., \& Boksenberg, A. 1983, in {\sl Quasars and
          Gravitational Lenses, 24th Lieg\`e International Colloquium}, p.518
\hangpara Sargent, W. L. W., Boksenberg, A. \& Steidel, C. C. 1988, ApJS, 68,
          539
\hangpara Verner, D. A., Tytler, D., \& Barthel, P. 1994, submitted to ApJL
\hangpara Williams, T. B. \& Schommer, R. A. 1993, ApJL, 419, L53

\bigskip
\vfill\eject
%
%
%
\centerline{FIGURE CAPTIONS}
\bigskip

\item{Fig. 1} Composite spectra at the expected positions of the C~IV
absorption lines in \Lya systems. The expected wavelengths of the C~IV
doublet lines are shown by the two vertical lines in each panel.
The number of spectral
segments included in the composite spectra depends on the wavelength:
65 at C~IV~$\lambda  1548$, 58 at C~IV~$\lambda  1550$,
38 at the blue end, and 27 at the red end.
In the top panel (a) the spectra are unweighted and the SNR=80.
In panel (b) the spectra are weighted by $N$(H~I) and the SNR=50.
The pixel size in these spectra is 0.025~\AA\ in the rest frame,
or 4.8 km~s$^{-1}$. The individual spectra have FWHM of about two pixels,
so the C~IV lines should be at least five pixels wide in the composite
spectra. The two lower
panels show the rest frame equivalent width for lines which are
five pixels wide. Panel (c) applies to spectrum (a) and panel (d) to spectrum
(b). Absorption lines are negative excursions in the spectra that are defined
to have  positive equivalent widths.

\bigskip

\item{Fig. 2} Histogram of the H~I column densities of the 65 \Lya
systems used in the composite spectrum. The bins are $10^{13}$~cm$^{-2}$
wide, the range is $0.85 \times 10^{13}$ -- $0.95 \times 10^{15}$,
and the mean is $1.0 \times 10^{14}$~cm$^{-2}$.

\vfill\eject
\bye